\documentclass[]{spie}  

\usepackage{amsmath,amsfonts,amssymb,aas_macros}
\usepackage{graphicx}
\usepackage[colorlinks=true, allcolors=blue]{hyperref}

\title{The GRAVITY fringe tracker: correlation between optical path residuals and atmospheric parameters}

\author[a,b]{S. Lacour}
\author[c]{R. Dembet}
\author[c]{R. Abuter}
\author[a]{P. Fedou}
\author[a]{G. Perrin}
\author[b]{F. Eisenhauer}
\author[d]{K. Perraut}
\author[e]{C. Straubmeier}
\author[f]{W. Brandner}
\author[g]{A. Amorin}
\author[ ]{the GRAVITY collaboration}
\affil[a]{LESIA/Observatoire de Paris/PSL Research University/CNRS, 92190 Meudon, France}
\affil[b]{Max Planck Institute for extraterrestrial Physics, D-85748 Garching, Germany}
\affil[c]{European Southern Observatory, 85748 Garching, Germany}
\affil[d]{Univ. Grenoble Alpes, CNRS, IPAG, 38000 Grenoble, France}
\affil[e]{1st Institute of Physics, University of Cologne, Zülpicher Straße 77, 50937 Cologne, Germany}
\affil[f]{Max Planck Institute for Astronomy, Königstuhl 17, 69117, Heidel- berg, Germany}
\affil[g]{Universidade de Lisboa - Faculdade de Ciências, Campo Grande,
1749-016 Lisboa, Portugal}

\authorinfo{Further author information: S.L. E-mail: sylvestre.lacour@obspm.fr}

\begin{document} 
\maketitle

\begin{abstract}
After the first year of observations with the GRAVITY fringe tracker, we compute correlations between the optical path residuals and atmospheric and astronomical parameters. The median residuals of the optical path residuals are 180\,nm on the ATs and 270\,nm on the UTs. The residuals are uncorrelated with the target magnitudes for Kmag below 5.5 on ATs (9 on UTs). The correlation with the coherence time is however extremely clear, with a drop-off in fringe tracking performance below 3\,ms.
\end{abstract}


\section{INTRODUCTION}
\label{sec:intro}  

The GRAVITY  \cite{2017A&A...602A..94G} instrument was installed in 2016 on Mount Cerro Paranal. It combines up to 4 beams of the ESO VLT interferometer. The GRAVITY fringe tracker (FT) is part of the instrument and corrects the atmospheric perturbations which blur the fringes and hinder long exposure times. The target of the FT was to maintain the optical path difference (OPD) below 300\,nm with a goal at 200\,nm.

The purpose of this short paper is to comment on the first year of fringe tracking (June 2016 - October 2017), to present the OPD residual levels, and to discuss the correlations between residuals and parameters such as target magnitude, seeing conditions, etc.

\section{Tracking accuracy}

\begin{figure} [ht]
   \begin{center}
   \begin{tabular}{c} 
   \includegraphics[height=8.5cm]{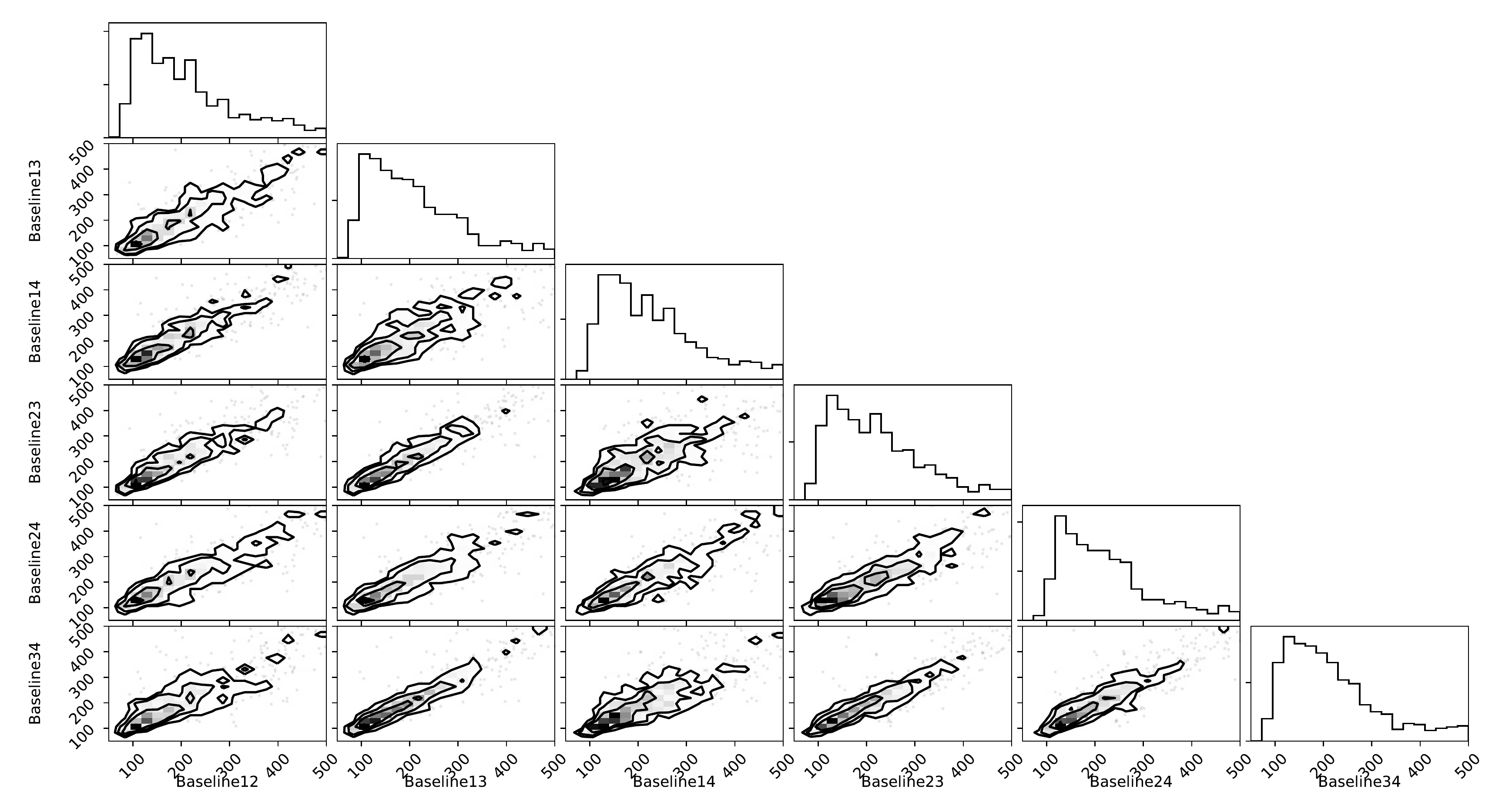}
   \end{tabular}
   \end{center}
   \caption[example] 
   { \label{fig:corrms} 
Correlation between the OPD residual error (in nanometers) observed on the different baselines. The residuals are obtained from the calibrators observed between June 2016 and October 2017. The median tracking residual is 180\,nm.
}
   \end{figure}

The four VLTI telescopes imply that 6 baselines are observable simultaneously. And indeed, GRAVITY tracks the fringes on all baselines simultaneously. Figure~\ref{fig:corrms} presents the histograms of the residuals on all six baselines. The data set consists in all the calibrators observed between June 2016 and October 2017. This corresponds to a total of 961 exposures.

The histograms reveal no differences between the various baselines. The median value is 180\,nm for the ATs and 270\,nm for the UTs. 

All the baseline residuals are highly correlated, meaning that the tracking accuracy is not baseline dependent, but caused by environmental parameters.

\section{Correlations with other parameters}

\begin{figure} [ht]
   \begin{center}
   \begin{tabular}{c} 
   \includegraphics[height=8.5cm]{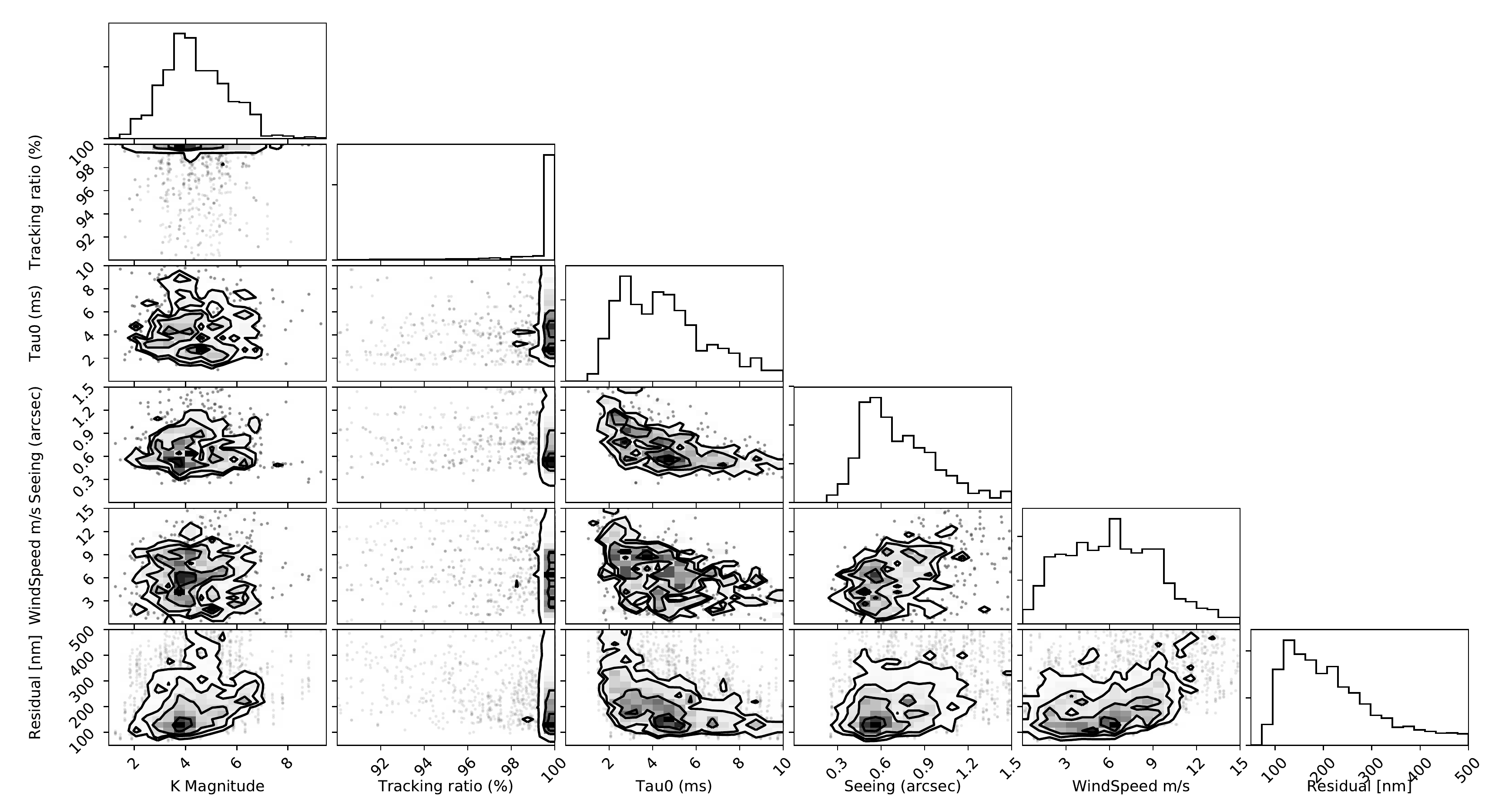}
   \end{tabular}
   \end{center}
   \caption[example] 
   { \label{fig:cor} 
Correlations between OPD residuals, K magnitude, tracking ratio, coherence time, seeing and wind speed (from the Paranal ASM system).
}
   \end{figure}

Figure~\ref{fig:cor} presents the OPD residuals as a function of tracking ratio, coherence time, seeing conditions and wind speed. The tracking ratio is the percentage of integration time during which the FT is in "phase tracking" mode. The FT is in this mode if the group delay residuals are below two fringes, ie $4.4\,\mu$m. The coherence time ($\tau_0$), seeing and wind speed values are taken from the Paranal astronomical site monitoring (ASM) system. $\tau_0$ is the coherence time in the visible. It corresponds to the time in milliseconds during which the variance of the phase (at $\lambda=500\,$nm) is below 1\ radian. The seeing is calculated at the same wavelength, and the wind speed  is the speed at the ground in m/s.

One may also look for correlations between fringe tracker residuals and target magnitude. In Figure~\ref{fig:cor}, the UT observations were shifted by 3.5\,mag to compare with AT observations. The faintest object was star TYC 5058-927-1 observed the 5th April 2017. The star has a magnitude Kmag$= 9.425$. The seeing conditions were excellent ($\tau_0\approx 12\,$ms), the integration time of the fringe tracker could be set to 10\,ms, and the tracking ratio was 100\% during the exposures. Overall the correlation with the magnitude follows two categories: below 5.5 the magnitude does not play a role in the fringe tracking performance. Above 5.5 the tracking accuracy decreases by around 70\,nm per magnitude -- up to 9.5, where the OPD residuals are of the order of 350\,nm.

The tracking ratio is always above 90\% -- the fringe tracker is mostly and ON/OFF system: either it works well or it does not work at all and the data does not appear in Figure~\ref{fig:cor}.

Of all parameters, the coherence time $\tau_0$ correlates best with the OPD residuals. Below 1.5\,ms, the fringe tracker does not work: there are almost no observations. Between 2 and 5\,ms, the median residuals are around 300\,nm. Above 6\,ms the residuals are almost always below 200\,nm.

Last the wind speed does not play a significant role below 7\,m/s. Above this value, the performance degrades to 200\,nm at 10\,m/s, up to 400\,nm at 15\,m/s. In the absence of wind, the performance also deteriorate due to convection at the telescope level.

\section{Summary}

\begin{itemize}
\item The performance of the Gravity fringe tracker is stable and optimal until a magnitude Kmag = 5.5 (9) for the Ats (UTs), and for a coherence time $\tau_0$ above 4\,ms. 
\item For fainter stars, residuals decrease by 70nm per magnitude (with a record faintest magnitude of 9.5). 
\item There is a threshold in coherence time ($\tau_0=3$\,ms) below which the OPD residuals increase considerably. 
\item The correlation between residuals and seeing is weak, hence seeing is not a good predictor of FT performances, while the coherence time is.
\end{itemize}

\acknowledgments 
 
 S.~Lacour acknowledges support from ERC starting grant No. 639248.
 
\bibliography{report,FTbib} 
\bibliographystyle{spiebib} 

\end{document}